\documentclass[preprint,12pt,authoryear]{elsarticle}

\usepackage{amssymb}
\usepackage{amsmath}
\usepackage{caption}
\usepackage{subcaption}
\usepackage{graphicx}
\usepackage{longtable}
\usepackage{multirow}
\usepackage{array}
\usepackage{pdflscape}
\usepackage{booktabs}
\usepackage[figuresright]{rotating}
\usepackage{multirow}
\usepackage{adjustbox}
\usepackage{lineno}

\usepackage[ruled,vlined,linesnumbered]{algorithm2e}
\newproof{pf}{Proof}
\usepackage{lineno}
\usepackage{float}
\usepackage{natbib}
\setcitestyle{numbers,square}
\usepackage{bm}
\newcommand{\R}{\mathbb{R}}
\newcommand{\E}{\mathbb{E}}
\DeclareMathOperator{\Var}{Var}
\DeclareMathOperator{\softmax}{softmax}

\DeclareMathOperator{\LeakyReLU}{LeakyReLU}
\DeclareMathOperator{\MaxPool}{MaxPool}
\DeclareMathOperator{\GAP}{GAP}
\DeclareMathOperator{\Dropout}{Dropout}
\DeclareMathOperator{\BN}{BN}

\usepackage{microtype}
\usepackage[none]{hyphenat}

\journal{Digital Signal Processing}

\begin{document}
\sloppy
\emergencystretch=3em

\begin{frontmatter}

\title{Integrating Vehicle Acoustic Data for Enhanced Urban Traffic Management: A Study on Speed Classification in Suzhou} 

\author[1]{Pengfei Fan}  
\author[1]{Yuli Zhang}  
\author[1]{Xinheng Wang*}  

\author[1]{Ruiyuan Jiang}  

\author[1]{Hankang Gu}  

\author[1]{Dongyao Jia}  

\author[2]{Shangbo Wang*}  


\affiliation[1]{organization={School of Advanced Technology},  
	addressline={Xi'an Jiaotong-Liverpool University},   
	city={Suzhou},  
	postcode={215123},   
	country={China}}  

\affiliation[2]{organization={School of Engineering and Informatics},  
	addressline={University of Sussex},   
	city={Brighton},  
	postcode={BN1 9RH},  
	country={UK}}

\begin{abstract}

This study presents and publicly releases the Suzhou Urban Road Acoustic Dataset (SZUR-Acoustic Dataset), which is accompanied by comprehensive data‐acquisition protocols and annotation guidelines to ensure transparency and reproducibility of the experimental workflow. To model the coupling between vehicular noise and driving speed, we propose a bimodal-feature-fusion deep convolutional neural network (BMCNN). During preprocessing, an adaptive denoising and normalization strategy is applied to suppress environmental background interference; in the network architecture, parallel branches extract Mel-frequency cepstral coefficients (MFCCs) and wavelet-packet energy features, which are subsequently fused via a cross-modal attention mechanism in the intermediate feature space to fully exploit time–frequency information. Experimental results demonstrate that BMCNN achieves a classification accuracy of 87.56\% on the SZUR-Acoustic Dataset and 96.28\% on the public IDMT-Traffic dataset. Ablation studies and robustness tests on the Suzhou dataset further validate the contributions of each module to performance improvement and overfitting mitigation. The proposed acoustics-based speed classification method can be integrated into smart-city traffic management systems for real-time noise monitoring and speed estimation, thereby optimizing traffic flow control, reducing roadside noise pollution, and supporting sustainable urban planning.
\end{abstract}




\end{frontmatter}




\section{Introduction}
Vehicle speed recognition is crucial in intelligent transportation systems \cite{balid1}. It optimizes traffic signal control and reduces congestion by monitoring and classifying vehicle speeds range in real time, while also identifying speeding or slow-moving vehicles to lower accident rates. Additionally, analyzing speed data helps predict traffic patterns and develop effective planning, providing information for navigation systems to assist drivers in choosing the best routes. Ultimately, intelligent traffic signal systems can dynamically adjust signal timings based on real-time speed data, enhancing the overall efficiency and safety of the transportation system \cite{asadi1}.

There are numerous methods for measuring the speed of moving vehicles, each with distinct advantages and limitations. Radar systems employing interferometric antenna apertures have proven to be effective and accurate for vehicle speed measurement \cite{Radar1,Radar2}. These radar systems are relatively low-cost, capable of operating under a wide range of weather conditions, and provide extensive coverage. However, they are susceptible to interference from external signals, and their accuracy in multi-lane speed measurement is limited due to challenges in distinguishing between vehicles in adjacent lanes.

To overcome these limitations, laser-based velocimeters, such as LiDAR systems, have been developed \cite{laser1,laser2}. LiDAR technology frequently scans the surrounding environment using laser pulses, enabling precise estimation of vehicle speed \cite{laser2}. These systems are particularly effective in scenarios requiring high accuracy and single-vehicle tracking. However, their performance is highly sensitive to environmental factors, such as adverse weather conditions, which can degrade their reliability and accuracy.

Video-based speed measurement is another widely used method. This approach leverages cameras to capture the trajectory of moving vehicles and employs advanced image processing techniques to calculate speed by analyzing positional changes over time \cite{Video1,Video2,Video3}. Video-based systems are advantageous in their ability to monitor multiple lanes simultaneously and provide additional information, such as vehicle type. However, these systems are significantly affected by weather conditions, such as poor lighting, heavy rain, or fog, which can impair image clarity and reduce recognition accuracy \cite{Video4,Video5}.

Satellite-based speed measurement, such as GPS systems, utilizes satellite positioning technology to record a vehicle's position at different time intervals and calculate its speed \cite{GPS1,GPS2,GPS3}. This method is often implemented using global navigation systems, combined with low-cost inertial measurement units for enhanced accuracy. While satellite-based systems are highly effective for real-time speed monitoring over large areas, they are limited in environments with signal obstruction, such as tunnels, urban canyons, or densely forested areas, where positioning accuracy may be compromised.

While radar, LiDAR, video-based, and satellite-based systems each offer unique advantages in vehicle speed measurement, their effectiveness depends on the specific application requirements and environmental conditions. Hybrid approaches that combine multiple technologies may help mitigate the limitations of individual systems and provide more robust and reliable speed measurement solutions. However, these speed measurement systems typically require significant equipment and capital investment. Moreover, traditional speed measurement systems demand substantial energy supplies to support their operation, further increasing their cost and environmental impact.

Additionally, the aforementioned systems are primarily designed to accurately measure the speed of individual vehicles. However, in some practical scenarios, it may not be necessary to measure the exact speed of each vehicle. Instead, the focus may be on assessing the overall traffic flow or the collective condition of vehicles on a road. For instance, measuring the average speed of vehicles on a road segment can help evaluate traffic congestion levels. In such cases, vehicle noise emissions can serve as an alternative means to estimate the overall speed range of vehicles \cite{tyagi1}.

Despite its potential, there are still limitations and deficiencies in classifying vehicle speed ranges through noise emissions. These include the lack of comprehensive datasets, underdeveloped classification methodologies, and the influence of environmental factors such as road conditions, traffic density, and background noise, which complicate accurate assessments. The main shortcomings of the current research are as follows:

\begin{itemize}
\item[1] The current availability of datasets on vehicle noise emissions is limited, lacking comprehensive records on vehicle speed, type, and operating conditions, which affects the accuracy of noise analysis and modeling.
\item[2] At present, there are few established methodologies for classifying noise emissions throughout the vehicle driving process, particularly in relation to the classification of vehicle driving speeds range. 
\end{itemize}

To address the aforementioned challenges, this study introduces a Suzhou Urban Road Acoustic Dataset (SZUR-Acoustic Dataset) of vehicle noise emissions recorded at various driving speeds. Additionally, the study identifies and analyzes the characteristic frequencies of vehicle noise emissions corresponding to different speed ranges, providing a comprehensive comparative analysis of various frequency selection methods. The primary contributions and innovations of this paper are summarized as follows:

\begin{itemize}
\item[1] We present a dataset recorded in an urban street in Suzhou which names Suzhou Urban Road Acoustic Dataset (SZUR-Acoustic Dataset), containing acoustic recordings and precise speed measurements of 4,822 vehicles at a key point. Each audio sample lasts two seconds, enabling effective noise analysis.
\item[2] We tense a frequency selection methodology that minimizes bandwidth usage while maintaining classification accuracy. This method supports optimal sampling rates for measurement devices and improves frequency selection in the classification process.
\item[3] We utilize a frequency-chosen two-feature network architecture (FC-TFNA) that employs a 2D Convolutional Neural Network (2D CNN) to extract features from Mel-Frequency Cepstral Coefficients (MFCC) and wavelet transforms. We concatenate these features at the flattening layer to enhance classification accuracy, followed by a Multi-Layer Perceptron (MLP) for precise speed classification across defined intervals.
\item[4] We validated this method on a public dataset for classifying vehicle speed ranges. Results demonstrate its potential for real-time traffic flow monitoring and congestion detection, providing efficient, data-driven support for urban traffic management.

\end{itemize}

This paper investigates several key areas. We review the state of the art in acoustic classification techniques within the transportation domain in Section 2, with particular emphasis on traffic noise identification, vehicle-type recognition, and road-condition detection, and we summarize the principal algorithms employed. We delineate the theoretical foundations of our work in Section 3, covering Mel-frequency cepstral coefficient (MFCC) extraction, wavelet transform analysis, and the bimodal-feature-fusion deep convolutional neural network (BMCNN) architecture, and we provide pseudo-code for its implementation. We introduce the Suzhou Urban Road Acoustic Dataset (SZUR-Acoustic Dataset) and the IDMT-Traffic dataset in Section 4, detailing their construction procedures, data characteristics, and key distinctions. We detail the evaluation metrics, loss functions, and hyperparameter configurations adopted for BMCNN in Section 5. We present our experimental results on the SZUR-Acoustic Dataset in Section 6, including classification accuracy, ablation studies, and analyses of model robustness and generalization capability. Finally, we conclude the paper in Section 7, summarizing our principal contributions and outlining directions for future research.

\section{Literature Review}\label{LR}
By examining the acoustic emissions generated during vehicle operation, researchers can classify or detect a range of operational parameters, vehicle states, and environmental conditions. This strategy relies on the acoustic signatures inherent in noise sources, thereby offering real-time insights into both the vehicle’s functional status and its surrounding context.

One notable example of such an approach is Bulatovi{\'c} et al. (2021), who presented a vehicle speed estimation method based on a single sensor and acoustic measurements, extracting features from Mel spectrograms. Their findings revealed an average velocity estimation error of 7.39 $\mathrm{km/h}$. When speeds were discretized in 10 $\mathrm{km/h}$ intervals, the accuracy reached 53.2\%, which increased to 93.4\% upon introducing an one-class tolerance shift. However, the authors underscored the limitations posed by a modest sample size and the overall accuracy \cite{LR2}.

Vehicle noise signals have also proven valuable in classifying vehicle types. For instance, Mohine et al. (2022) developed a hybrid deep learning architecture integrating an one-dimensional convolutional neural network (CNN) with bidirectional long short-term memory (BiLSTM) network. This model effectively identified two-wheelers as well as low, medium, and heavy vehicles, even in noisy environments, by automatically extracting acoustic features while capturing temporal dependencies. With a classification accuracy of 96\% on the SITEX02 dataset, the proposed method demonstrated considerable robustness and efficacy \cite{LR3}.

A further extension of acoustic-based classification emerges in urban noise analysis. Tran et al. (2020) introduced SirenNet, a CNN-based model composed of two parallel network streams: WaveNet, which processes raw waveform data, and MLNet, which utilizes MFCC and log-Mel spectrograms. Their experiments indicated that raw waveform features significantly complement MFCC and log-Mel representations, culminating in a 98.24\% accuracy for alarm sound detection. Notably, the system maintained a 96.89\% accuracy rate even for 0.25-second audio samples, underscoring its resilience to variations in input length \cite{LR4}.

Lastly, Yoo et al. (2022) highlighted the utility of tire–pavement interaction noise (TPIN) for classifying road surface conditions. By employing a CNN, their framework discriminated effectively between snow-covered and non-snow-covered roads, thereby offering a preliminary assessment of frictional properties. This underscores the broader significance of acoustic signals in elucidating critical information about road infrastructure, vehicle safety, and operational conditions \cite{LR5}.

An acoustic signal processing, the extraction of speech features constitutes a fundamental and pivotal step, as it significantly influences the performance of downstream tasks such as speech recognition, speaker identification, and emotion analysis. Over the years, numerous methodologies have been developed for speech feature extraction, each designed to capture distinct characteristics of the speech signal. This paper presents a comprehensive review of several widely adopted approaches, emphasizing their underlying principles, strengths, and practical applications.

One of the most extensively utilized techniques is the MFCC, which transforms the speech spectrum onto the Mel scale. This scale approximates the nonlinear frequency perception of the human auditory system, enabling the extraction of cepstral coefficients that effectively represent the spectral envelope \cite{LR6.01,MFCCM,LR6.03,chen2}. MFCC is particularly advantageous due to their ability to capture perceptually relevant features of speech, making them a cornerstone in many speech processing systems. Mishra et al. (2025) employed MFCC to extract acoustic features, which were further enhanced by entropy-based features derived from the Multi-Resolution Hilbert Transform (MRHEF) to identify speech emotion characteristics. A deep neural network classifier was subsequently utilized to classify emotions, demonstrating the effectiveness of this approach in achieving high accuracy \cite{LR6}.

Linear Predictive Cepstral Coefficients (LPCC), derived from Linear Predictive Coding (LPC), model the vocal tract as an all-pole filter. These cepstral coefficients are computed from LPC parameters and are particularly effective in capturing the resonant properties of speech signals \cite{LR7.01}. LPCC is known for their computational efficiency and robustness in representing speech characteristics, particularly in low-resource settings. In a comparative study by Hariharan et al. (2012), the performance of MFCC and LPCC were evaluated for detecting stuttering events. The findings indicated that both MFCC and LPCC are suitable for stuttering detection, with LPCC demonstrating slight performance advantage over MFCC due to its ability to better model the resonances of the vocal tract \cite{MFCC}.

Beyond traditional techniques, the wavelet transform has emerged as a powerful method for analyzing signals in both the time and frequency domains simultaneously, making it particularly suitable for processing non-stationary signals such as speech. Unlike Fourier-based methods, the wavelet transform provides multi-resolution analysis, enabling the effective extraction of both global and local signal characteristics \cite{LR8,LR9,LR10}. Anuragi et al. (2022) proposed an automatic cross-subject emotion recognition framework based on EEG signals, utilizing the empirical wavelet transform method derived from Fourier-Bessel series expansion (FBSE-EWT). Their study demonstrated the superior performance of this approach in classifying human emotions, highlighting its potential not only in speech analysis but also in broader applications such as biomedical signal processing \cite{LR11}.

The choice of feature extraction method depends on the specific application and the nature of the speech signal being analyzed. While MFCC and LPCC remain widely used due to their simplicity and effectiveness, advanced techniques like wavelet transforms offer greater flexibility and precision in capturing complex signal dynamics, particularly for non-stationary and multi-dimensional data. These methods continue to evolve, driven by advancements in computational power and the growing demands of modern acoustic analytics.

In the broader context of acoustic analytics, research on classifying physical objects based on sound features has demonstrated notable advancements across a range of domains. Numerous classification algorithms, spanning traditional machine learning techniques to more recent deep learning approaches, have been employed to enhance performance and reliability in these classification tasks \cite{LR12}. These advancements have been driven by the increasing availability of diverse datasets and the growing computational power that enables the application of more complex models.

In the process of handling simple classification problems, traditional machine learning methods have proven effective \cite{LR12,LR19}. These methods primarily include K-Nearest Neighbors (KNN) \cite{LR13}, Gaussian Mixture Model (GMM) \cite{LR14}, Support Vector Machine (SVM) \cite{LR15}, and Hidden Markov Model (HMM) \cite{LR16,LR17}, and decision trees \cite{LR18}. However, with the evolution of data-driven approaches, supervised learning methods have become increasingly prominent in acoustic classification tasks. CNN and their optimized frameworks are widely utilized in this domain. For instance, Kwon (2021) employed a 1-dimensional dilated CNN end-to-end model for human speech emotion recognition, achieving recognition accuracies of 73\% and 90\% on the IEMOCAP and EMO-DB datasets, respectively \cite{LR21}. Furthermore, combining CNNs with other models has shown significant potential in improving classification performance. Ahmend et al. (2023) proposed an integrated model of 1D-CNN-LSTM-GRU for speech emotion recognition, achieving accuracy rates of 99.46\% on TESS, 95.42\% on EMO-DB, 95.62\% on RAVDESS, 93.22\% on SAVEE, and 90.47\% on the CREMA-D dataset \cite{LR22}. In addition, advancements in CNN architectures, such as 2D-CNNs, have further expanded their applicability. Zhao et al. (2021) combined 2D CNNs with self-attention expanded residual networks for human speech emotion recognition, achieving a weighted accuracy (WA) of 73.1\% and an unweighted accuracy (UA) of 66.3\% on the IEMOCAP dataset, and a UA of 41.1\% on the FAU-AEC dataset \cite{LR23}.

Beyond CNNs, recurrent neural networks such as Long Short-Term Memory (LSTM) networks \cite{LR24,LR25,LR28} and Gated Recurrent Units (GRUs) \cite{LR26,LR27} have also demonstrated broad applicability in acoustic classification tasks. Lu et al. (2024) introduced a novel late fusion framework, the Multimodal Residual Speaker-LSTM Network (MRSLN), designed to address the over-smoothing issue in deep LSTM networks. By incorporating speaker-specific contextual information, the MRSLN effectively captures both inter-speaker and intra-speaker interactions. Extensive evaluations on the IEMOCAP and MELD datasets revealed that MRSLN not only achieved superior classification accuracy but also offered enhanced computational efficiency, outperforming existing state-of-the-art (SOTA) models. This work underscores the potential of MRSLN as a robust and efficient solution for unimodal learning tasks \cite{LR29}. 

\section{Methodology}\label{M}
In this section, we will explore the fundamental principles of MFCC, wavelet transform, 2D CNN, and MLP adopted in this study. These techniques play a key role in feature extraction and classification. Understanding these theoretical concepts will help us better comprehend the design and optimization of the models.
\subsection{Mel-frequency Cepstral Coefficients}
MFCC is a widely used feature extraction technique in speech and audio processing, involving several key steps. The process begins with preprocessing the audio signal, which includes noise reduction, framing, and windowing to improve signal quality. Next, the Fast Fourier Transform (FFT) converts the time-domain signal into the frequency domain, allowing for spectral analysis. Following this, a set of Mel frequency filters is applied to the FFT results, simulating the auditory characteristics of the human ear and aligning the frequency representation with human perception of pitch. The logarithm of each filter's output is then computed to compress the dynamic range, enhancing robustness against amplitude variations. The mapping from linear frequency to Mel-frequency is mathematically represented by the following formula \cite{MFCC,MFCCM}:
\begin{equation}  
f_m = 2595 \cdot \log_{10}\left(1 + \frac{f}{700}\right) 
\end{equation}
where $f$ is frequency (Hz), $f_m$ is Mel frequency. 
\subsection{Wavelet Transform}
Wavelet transform is a sophisticated mathematical framework employed for the analysis and processing of signals and data. Unlike traditional Fourier transform, which represents signals through sine and cosine functions, wavelet transform offers a time-frequency representation that enables the examination of signals across various scales and resolutions. This characteristic makes it particularly advantageous for analyzing non-stationary signals, where the frequency content exhibits temporal variability \cite{Wavelet1,Wavelet2}.

In the context of wavelet transform, the Coif1 wavelet is regarded as an effective tool for conducting multi-resolution analysis of signals. By decomposing a signal into its constituent frequency components, the Coif1 wavelet facilitates the simultaneous extraction of both low-frequency and high-frequency information. When using the Coif1 wavelet for transformation, the signal is filtered through the scaling function, yielding a series of approximation coefficients and detail coefficients that encapsulate the characteristics of the signal at different scales, thereby enhancing the potential for subsequent analysis and processing. The Coif1 wavelet is widely recognized for its applicability across various domains, including image compression, noise reduction, and feature extraction. Its favorable mathematical properties and adaptability contribute to its effectiveness in processing non-stationary signals, allowing for precise capture of instantaneous changes and local features within the signal.

The scaling function is utilized to describe the low-frequency components of a signal. In wavelet transforms, it plays a crucial role in smoothing the signal. The function of scaling function is:
\begin{equation}  
\phi(t) = \sum_{n=0}^{N-1} h(n) \cdot \phi(2t - n)  
\label{eq:scaling_function}  
\end{equation}
where $t$ denotes the time or spatial variable, representing the input value of the signal, $h(n)$ are the coefficients of the scaling function, which determine the shape and characteristics of the scaling function. These coefficients are computed through specific algorithms, such as orthogonality conditions, $N$ indicates the length of the scaling function, representing the number of coefficients. The mother wavelet function is employed to describe the high-frequency components of a signal. In wavelet transforms, it is utilized to capture the instantaneous variations and details of the signal. The formula of mother funnction is:
\begin{equation}  
\psi(t) = \sum_{n=0}^{N-1} g(n) \cdot \phi(2t - n)  
\label{eq:wavelet_function}  
\end{equation}
where $g(n)$ are the coefficients of the mother wavelet, which determine the shape and characteristics of the wavelet function. 

\subsection{Balanced Multi-Modal CNN (BMCNN)}

\subsubsection{Mathematical Formulation of Data Preprocessing}
The Z-score standardization constitutes a linear transformation technique that maps raw data into a standard normal distribution space, thereby eliminating scale disparities across different feature dimensions and ensuring balanced contributions of each feature to model training. This transformation is grounded in the statistical principles of the Central Limit Theorem.

\begin{equation}
\mathbf{X}_{\text{standardized}} = \frac{\mathbf{X} - {\mu}}{{\sigma} + \epsilon}
\end{equation}
where: $\mathbf{X} \in \R^{n \times d}$ denotes the raw input feature matrix, with $n$ representing the number of samples and $d$ the feature dimensionality. $\bm{\mu} \in \R^d$ represents the feature mean vector, defined as $\bm{\mu} = \E[\mathbf{X}] = \frac{1}{n}\sum_{i=1}^{n} \mathbf{X}_i$. $\bm{\sigma} \in \R^d$ denotes the feature standard deviation vector, computed as $\bm{\sigma} = \sqrt{\Var[\mathbf{X}]} = \sqrt{\frac{1}{n}\sum_{i=1}^{n} (\mathbf{X}_i - \bm{\mu})^2}$. $\epsilon$ represents a numerical stability constant, $\epsilon = 10^{-8}$, employed to prevent division by zero. The outlier detection and clipping technique based on the $3\sigma$ criterion leverages the statistical properties of normal distributions, wherein 99.7\% of data points fall within three standard deviations. This approach enhances model robustness against noise by truncating extreme values.

\begin{equation}
\mathbf{X}_{\text{clipped}} = \max(-3, \min(3, \mathbf{X}_{\text{standardized}}))
\end{equation}
Min-Max normalization represents a shape-preserving linear transformation that maps data to the $[0,1]$ interval, mitigating the impact of numerical ranges on activation function gradient propagation.

\begin{equation}
\mathbf{X}_{\text{normalized}} = \frac{\mathbf{X}_{\text{clipped}} - \mathbf{X}_{\min}}{\mathbf{X}_{\max} - \mathbf{X}_{\min} + \epsilon}
\end{equation}
where: $\mathbf{X}_{\min} = \min(\mathbf{X}_{\text{clipped}})$ denotes the minimum value of clipped data. $\mathbf{X}_{\max} = \max(\mathbf{X}_{\text{clipped}})$ represents the maximum value of clipped data

\subsubsection{Mathematical Theory of Convolutional Neural Networks}

The convolution operation constitutes the fundamental mathematical operation in CNNs, rooted in the convolution theorem of linear system theory. In deep learning contexts, convolution operations achieve translation invariance and local connectivity while substantially reducing model complexity through parameter sharing.
\begin{equation}
\mathbf{Y}[i,j] = (\mathbf{X} * \mathbf{K})[i,j] = \sum_{m=0}^{M-1} \sum_{n=0}^{N-1} \mathbf{X}[i+m, j+n] \cdot \mathbf{K}[m,n] + b
\end{equation}
where: $\mathbf{Y} \in \R^{H' \times W'}$ denotes the output feature map. $\mathbf{X} \in \R^{H \times W}$ represents the input feature map. $\mathbf{K} \in \R^{M \times N}$ signifies the convolution kernel (filter) weight matrix. $b \in \R$ denotes the bias parameter. $*$ represents the convolution operator. $(i,j)$ indicates spatial coordinates in the output feature map. $(m,n)$ denotes local coordinates within the convolution kernel. Batch Normalization addresses the Internal Covariate Shift problem by normalizing activations within each mini-batch, thereby accelerating training convergence and enhancing model generalization performance.

\begin{equation}
\BN(\mathbf{x}) = \bm{\gamma} \odot \left(\frac{\mathbf{x} - \bm{\mu}_B}{\sqrt{\bm{\sigma}_B^2 + \epsilon}}\right) + \bm{\beta}
\end{equation}
where: $\bm{\mu}_B = \frac{1}{m}\sum_{i=1}^{m} \mathbf{x}_i$ represents the batch sample mean. $\bm{\sigma}_B^2 = \frac{1}{m}\sum_{i=1}^{m} (\mathbf{x}_i - \bm{\mu}_B)^2$ denotes the batch sample variance. $\bm{\gamma}, \bm{\beta} \in \R^d$ are learnable scale and shift parameters. $\odot$ denotes element-wise multiplication (Hadamard product). $m$ represents the batch size. The Leaky ReLU represents an enhancement of the traditional ReLU function, mitigating the ``dying ReLU'' problem by introducing a small non-zero gradient for negative inputs, thereby maintaining gradient flow throughout the network.

\begin{equation}
\LeakyReLU(x) = \max(\alpha x, x) = \begin{cases} 
x, & \text{if } x \geq 0 \\
\alpha x, & \text{if } x < 0 
\end{cases}
\end{equation}
where: $\alpha$ denotes the negative slope parameter, typically set as $\alpha = 0.1$. Max pooling constitutes a non-linear downsampling operation that preserves salient features while reducing spatial dimensions of feature maps, providing translation invariance and computational efficiency.

\begin{equation}
\MaxPool(\mathbf{X})[i,j] = \max_{0 \leq p < k, 0 \leq q < k} \mathbf{X}[i \cdot s + p, j \cdot s + q]
\end{equation}
where: $s$ denotes the stride parameter. $k$ represents the pooling window size. $(p,q)$ indicates relative coordinates within the pooling window. Global Average Pooling (GAP) transforms two-dimensional feature maps into one-dimensional vectors by computing the spatial mean, offering superior regularization effects and reduced parameter count compared to fully connected layers.

\begin{equation}
\GAP(\mathbf{X}) = \frac{1}{H \cdot W} \sum_{i=1}^{H} \sum_{j=1}^{W} \mathbf{X}[i,j]
\end{equation}
where: $H$ denotes the feature map height. $W$ represents the feature map width. Dropout represents a stochastic regularization technique that prevents overfitting by randomly deactivating neural outputs during training, grounded in ensemble learning theory.

\begin{equation}
\Dropout(x) = \begin{cases}
\frac{x}{1-p}, & \text{with probability } (1-p) \\
0, & \text{with probability } p
\end{cases}
\end{equation}
where: $p \in [0,1]$ denotes the dropout probability.

\subsubsection{Regularization Theory and Loss Function Construction}

L2 regularization incorporates an L2 norm penalty term on weight parameters into the loss function, controlling model complexity based on Occam's razor principle to prevent overfitting.
\begin{equation}
\Omega(\bm{\theta}) = \lambda \sum_{l=1}^{L} \|\mathbf{W}^{(l)}\|_F^2 = \lambda \sum_{l=1}^{L} \sum_{i,j} (W_{i,j}^{(l)})^2
\end{equation}
where: $\lambda > 0$ represents the regularization strength hyperparameter, $\lambda = 5 \times 10^{-3}$. $\mathbf{W}^{(l)}$ denotes the weight matrix of layer $l$. $\|\cdot\|_F$ represents the Frobenius norm. $L$ indicates the number of network layers. The total loss function combines empirical risk and structural risk, achieving the bias-variance tradeoff.
\begin{equation}
\mathcal{L}_{\text{total}}(\bm{\theta}) = \mathcal{L}_{\text{empirical}}(\bm{\theta}) + \Omega(\bm{\theta})
\end{equation}
where: $\mathcal{L}_{\text{empirical}}(\bm{\theta})$ represents the empirical loss (e.g., cross-entropy loss). $\Omega(\bm{\theta})$ denotes the regularization term.

\subsubsection{Optimization Algorithms and Loss Functions}

The cross-entropy loss, rooted in information theory, measures the Kullback-Leibler divergence between predicted probability distributions and ground truth distributions, serving as the standard loss function for multi-class classification problems.
\begin{equation}
\mathcal{L}_{\text{CE}} = -\sum_{i=1}^{N} \sum_{c=1}^{C} y_{i,c} \log(\hat{y}_{i,c} + \epsilon)
\end{equation}
where: $N$ denotes the number of samples. $C$ represents the number of classes. $y_{i,c}$ indicates the one-hot encoded ground truth label. $\hat{y}_{i,c}$ represents the model's predicted probability for class $c$. The Adam optimizer combines advantages of momentum-based gradient descent and RMSprop, achieving adaptive learning rate adjustment through exponential moving averages of first and second moments of gradients.
\begin{align}
\mathbf{m}_t &= \beta_1 \mathbf{m}_{t-1} + (1-\beta_1)\mathbf{g}_t \\
\mathbf{v}_t &= \beta_2 \mathbf{v}_{t-1} + (1-\beta_2)\mathbf{g}_t^2 \\
\hat{\mathbf{m}}_t &= \frac{\mathbf{m}_t}{1-\beta_1^t} \\
\hat{\mathbf{v}}_t &= \frac{\mathbf{v}_t}{1-\beta_2^t} \\
\bm{\theta}_{t+1} &= \bm{\theta}_t - \alpha \cdot \frac{\hat{\mathbf{m}}_t}{\sqrt{\hat{\mathbf{v}}_t} + \epsilon}
\end{align}
where: $\alpha$ denotes the initial learning rate, $\alpha = 5 \times 10^{-4}$. $\beta_1, \beta_2$ represent exponential decay rates, $\beta_1 = 0.9, \beta_2 = 0.999$. $\mathbf{g}_t = \nabla_{\bm{\theta}}\mathcal{L}(\bm{\theta}_t)$ indicates the gradient vector at step $t$. $\mathbf{m}_t$ represents the first moment estimate (momentum term). $\mathbf{v}_t$ denotes the second moment estimate (adaptive term). $t$ indicates the time step index. $\bm{\theta}$ represents the model parameter vector. The validation loss-based learning rate scheduling strategy dynamically reduces the learning rate when validation performance plateaus, preventing convergence to local optima.
\begin{equation}
\alpha_{\text{new}} = \alpha_{\text{old}} \times \gamma, \quad \text{if} \quad \mathcal{L}_{\text{val}}^{(t)} \geq \mathcal{L}_{\text{val}}^{(t-\text{patience})}
\end{equation}
where: $\gamma$ denotes the decay factor, $\gamma = 0.8$. $\text{patience}$ represents the tolerance epochs. $\mathcal{L}_{\text{val}}$ indicates the validation set loss.

\subsubsection{Multimodal Feature Fusion Architecture}
Multimodal feature fusion integrates information from different modalities through vector concatenation in the feature space, achieving synergistic representation of complementary features.
\begin{equation}
\mathbf{z} = [\mathbf{f}_{\text{MFCC}}; \mathbf{f}_{\text{Wavelet}}] \in \R^{d_1 + d_2}
\end{equation}
where: $\mathbf{f}_{\text{MFCC}} \in \R^{d_1}$ represents the feature vector extracted from the MFCC branch. $\mathbf{f}_{\text{Wavelet}} \in \R^{d_2}$ denotes the feature vector extracted from the wavelet branch. $[\cdot;\cdot]$ indicates the vector concatenation operation. Fully connected layers implement linear transformations that map high-dimensional features to target spaces, constituting fundamental computational units in neural networks.
\begin{equation}
\mathbf{h}^{(l+1)} = \sigma(\mathbf{W}^{(l)}\mathbf{h}^{(l)} + \mathbf{b}^{(l)})
\end{equation}
where: $\mathbf{W}^{(l)} \in \R^{n_{l+1} \times n_l}$ denotes the weight matrix of layer $l$. $\mathbf{b}^{(l)} \in \R^{n_{l+1}}$ represents the bias vector. $\sigma(\cdot)$ indicates the non-linear activation function. $\mathbf{h}^{(l)}$ denotes the activation values of layer $l$. The Softmax function maps real-valued vectors to probability distributions, ensuring outputs satisfy probability axioms: non-negativity and normalization constraints.
\begin{equation}
p(y=c|\mathbf{x}) = \frac{\exp(z_c)}{\sum_{j=1}^{C} \exp(z_j)} = \softmax(z_c)
\end{equation}
where: $z_c$ represents the logit output for class $c$. $C$ denotes the total number of classes. $p(y=c|\mathbf{x})$ indicates the predicted probability of class $c$ given input $\mathbf{x}$.
\subsection{Pseudocode}
\begin{figure}[ht]
	\centering
	\includegraphics[width=1\textwidth]{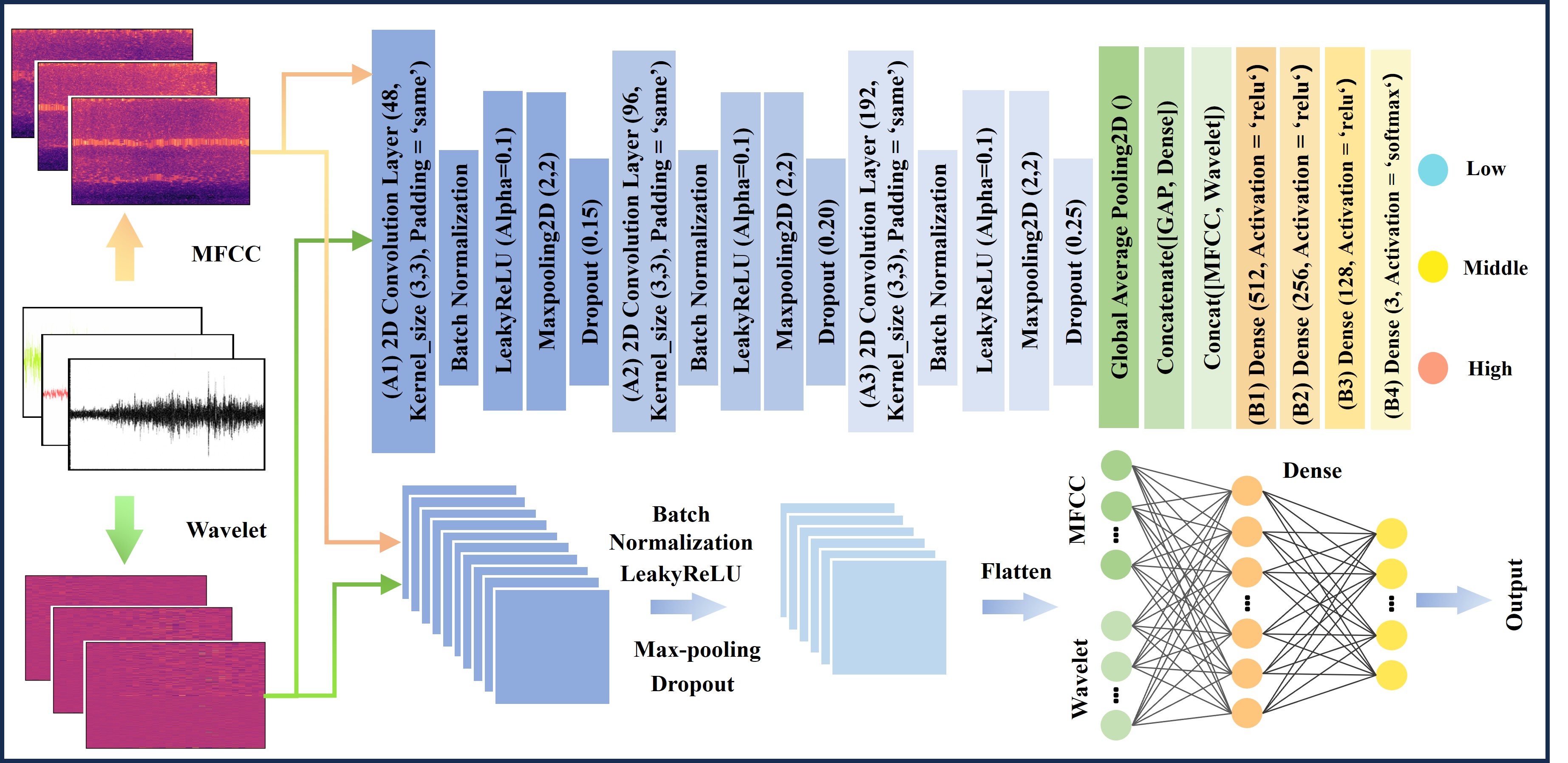}     
        \caption{Structure of bimodal-feature-fusion deep convolutional neural network (BMCNN).}
	\label{Structure}
\end{figure}
In this approach, we first load and initialize two modalities of features—MFCC (denoted \(X_M\)) and wavelet‐transform features (denoted \(X_W\))—along with their corresponding speed labels \(y\) from pre‐saved \texttt{.mat} files, which show in Figure \ref{Structure} and Algorithm 1. For each modality, we compute the mean and standard deviation to apply Z‐score normalization, then clip the normalized values to \([-3,3]\) and linearly rescale them to \([0,1]\) to improve training stability. The normalized feature maps are reshaped into tensors suitable for 2D convolutional input, and the dataset is partitioned into training, validation, and test splits by stratified sampling on the speed labels to ensure an even distribution of speed categories across all splits. Architecturally, each modality branch passes through three convolutional modules with channel sizes of 48, 96, and 192, respectively. Each module follows the sequence  
$\mathrm{Conv2D}\;\to\;\mathrm{BatchNorm}\;\to\;\mathrm{LeakyReLU}\;\to\;\mathrm{Dropout}\;\to\;\mathrm{MaxPool}$. Their outputs are fed into a Global Average Pooling layer and then a small fully connected layer to produce flattened feature vectors. The two modality‐specific vectors are concatenated into a single fused representation, which is subsequently passed through two fully connected layers—the first with 128 units and the second with a number of units equal to the speed category count—followed by a softmax activation to yield class probabilities. We train the network using the Adam optimizer (learning rate \(5\times10^{-4}\)) and sparse categorical cross‐entropy loss. After each epoch, a ReduceLROnPlateau scheduler adapts the learning rate, and the dropout rate is progressively increased to mitigate overfitting, with early stopping based on validation performance. Finally, we evaluate the model’s classification accuracy on the test set and report additional metrics—mean squared error (MSE), mean absolute error (MAE), and Cohen’s \(\kappa\) coefficient—to assess the consistency of the predicted speed labels.

\begin{algorithm}[H]
\caption{Training and Evaluation of BMCNN Algorithm}
\KwIn{$X_M$: MFCC features, $X_W$: Wavelet transform features, $y$: Speed Labels}
\KwOut{Multi-modal fusion CNN model $M$}

\textbf{Initialization:} Load $X_M$, $X_W$, $y$ from .mat files\;

\textbf{Preprocessing:} \For{each $X \in \{X_M, X_W\}$}{
    $\mu = \text{mean}(X, \text{axis}=0)$, $\sigma = \text{std}(X, \text{axis}=0)$\;
    $\tilde{X} = \text{clip}\left(\frac{X - \mu}{\sigma + \epsilon}, -3, 3\right)$\;
    $\hat{X} = \frac{\tilde{X} - \min(\tilde{X})}{\max(\tilde{X}) - \min(\tilde{X}) + \epsilon}$\;
}

\textbf{Reshape:} $T_M \in \mathbb{R}^{N \times T_M \times F_M \times 1}$, $T_W \in \mathbb{R}^{N \times T_W \times F_W \times 1}$\;

\textbf{Split:} $T_M^{tr}$, $T_M^{te}$, $T_W^{tr}$, $T_W^{te}$, $y^{tr}$, $y^{te}$ with stratified sampling\;
\For{each modality $i \in \{M, W\}$}{
    Conv2D$(48, (3,3)) \rightarrow$ BatchNorm $\rightarrow$ LeakyReLU $\rightarrow$ Dropout$(0.15) \rightarrow$ MaxPool2D\;
    Conv2D$(96, (3,3)) \rightarrow$ BatchNorm $\rightarrow$ LeakyReLU $\rightarrow$ Dropout$(0.2) \rightarrow$ MaxPool2D\;
    Conv2D$(192, (3,3)) \rightarrow$ BatchNorm $\rightarrow$ LeakyReLU $\rightarrow$ Dropout$(0.25) \rightarrow$ MaxPool2D\;
    $h_{gap} = \text{GlobalAvgPool2D}()$, $h_{flat} = \text{Dense}_{256}(\text{Flatten}())$\;
    $h_i = \text{Concatenate}([h_{gap}, h_{flat}])$\;
}
$h = \text{Concatenate}([h_M, h_W])$\;
$h \rightarrow \text{Dense}_{512} \rightarrow \text{Dense}_{256} \rightarrow \text{Dense}_{128} \rightarrow \text{Dense}_{C}$\;
Output: $\hat{y} = \text{Softmax}(h)$\;

\textbf{Compile:} $M$ with optimizer $=$ Adam$(lr=0.0005)$, loss $=$ sparse\_categorical\_crossentropy\;

\For{$e = 1$ \KwTo $250$}{
    Train $M$ on $[T_M^{tr}, T_W^{tr}]$, $y^{tr}$ with validation\_split $= 0.2$\;
    Apply ReduceLROnPlateau(factor$=0.8$, patience$=8$)\;
    Monitor training progress and adjust dropout progressively\;
}
$\hat{y}_{proba} = M([T_M^{te}, T_W^{te}])$, $\hat{y} = \arg\max(\hat{y}_{proba})$\;
$R = \text{accuracy}(y^{te}, \hat{y})$\;
$\text{MSE} = \frac{1}{N} \sum_{i=1}^{N} (y_i^{te} - \hat{y}_i)^2$, $\text{MAE} = \frac{1}{N} \sum_{i=1}^{N} |y_i^{te} - \hat{y}_i|$\;
$\kappa = \text{cohen\_kappa\_score}(y^{te}, \hat{y})$\;

\end{algorithm}
\section{Introduction of Dataset}\label{ID}
This section introduces two datasets used in this study, one is IDMT-Traffic dataset \cite{Dataset} and the other SZUR-Acoustic Dataset. The SZUR-Acoustic Dataset will be described in this section. IDMT-traffic is an open benchmark dataset. 

\subsection{SZUR-Acoustic Dataset}
\begin{figure}[ht]
	\centering
	\includegraphics[width=1\textwidth]{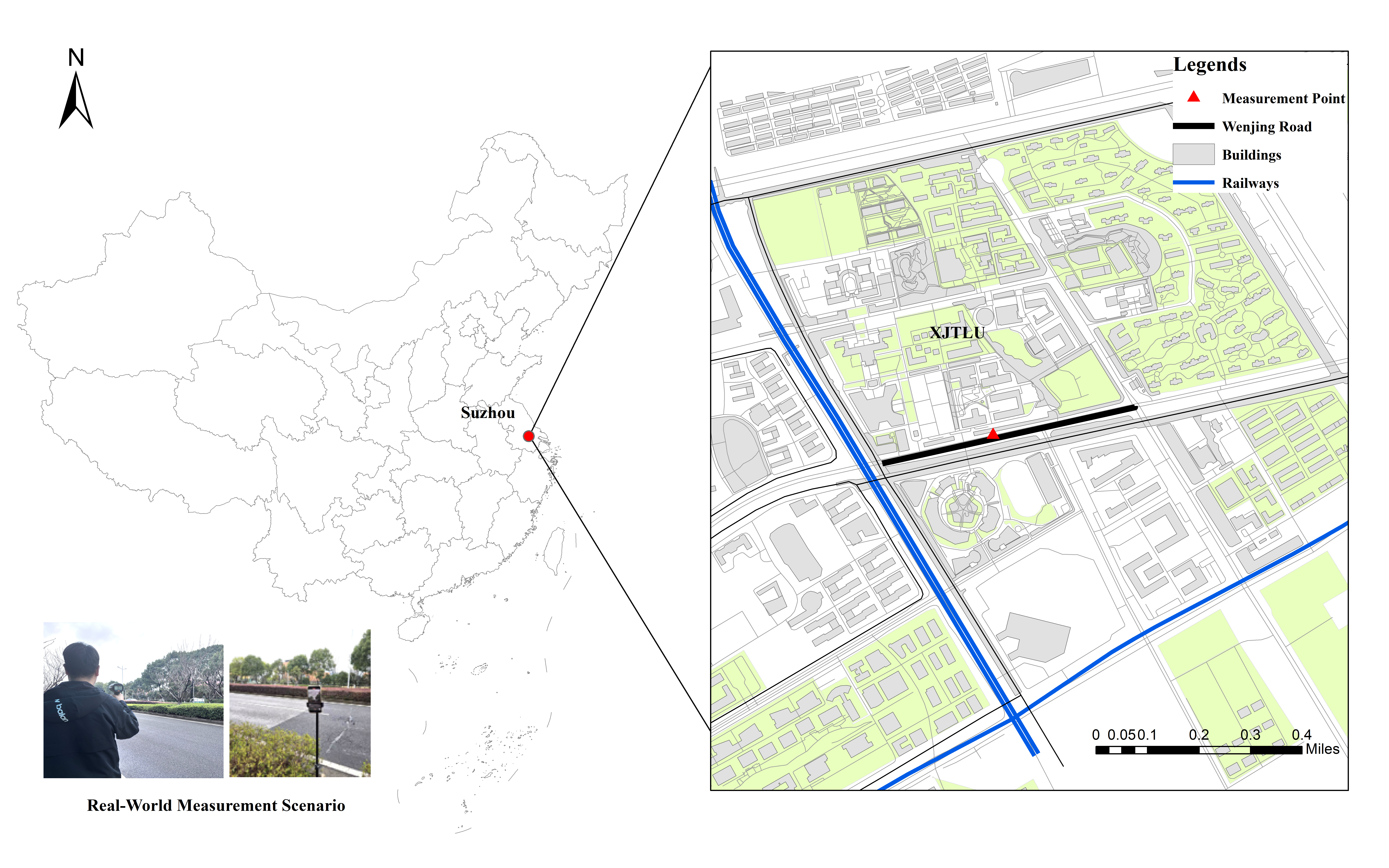}     
        \caption{Measurement Point Location Map.}
	\label{MPL}
\end{figure}

The dataset was collected in Wenjing Road, Suzhou, Jiangsu Province, China, with the intention of analyzing environmental noise levels in an urban area, which show in Figure \ref{MPL}. In this study, acoustic data was recorded using a cellphone with high-definition recording capabilities instead of a professional microphone, which enables more accessible and cost-effective data collection while still capturing relevant acoustic information.
Figure \ref{MPL} presents a schematic representation of the measurement location. The selected road features a dual-lane configuration on one side, complemented by a green belt that effectively separates the opposing traffic flows. This specific arrangement was deliberately designed to mitigate interference from noise originating from the opposite side of the road, thereby optimizing the integrity and quality of the recorded acoustic data. The experimental setup entailed the strategic placement of a cellphone on one side of the road, ensuring that it was sufficiently shielded from potential cross-traffic noise, thus enhancing the reliability of the sound recordings.

In selecting the roadway for this study, it is a typical urban road that the speed limit is set at 60  $\mathrm{km/h}$, and the roadway typifies an urban environment where no minimum speed limit is enforced. The noise recording point and speed test point were deliberately set more than 100 meters away from the nearest traffic signal, which effectively reduces the impact of sudden vehicle deceleration or acceleration typically occurring near intersections, ensuring that vehicles maintain a constant speed as they pass the measurement points. Data collection was conducted during nighttime or midday to further reduce the likelihood of confounding factors at the measurement site, such as a high volume of non-motorized vehicles or other disturbances. Consequently, it can be reasonably assumed that the speed of vehicles passing the measurement point remains relatively constant.

The dataset gathered in Suzhou was collected during clear weather and low wind speeds, which provided ideal conditions for accurate measurements. However, it's necessary to note that the analysis excluded sounds from insects and noise from two-wheeled vehicles on nearby non-motorized lanes, which could have influenced the overall acoustic environment during the measurement period.

To conduct the measurements, we implemented a strategy of measuring for one hour daily over 30 days. This method allowed us to circumvent unfavorable weather and extreme traffic conditions, thereby enhancing the reliability of the data collected.

To further supplement the acoustic recordings, a speed measurement point was established 50 meters away from the sound recording location. This distance is deliberately chosen to enable the speed measurement equipment to operate effectively and to minimize speed measurement errors caused by measuring distances that are too short. Additionally, the recording device was positioned at a height of 1.5 meters above ground level to more accurately capture the noise of passing vehicles. 

\begin{figure}[ht]
	\centering
	\includegraphics[width=0.7\textwidth]{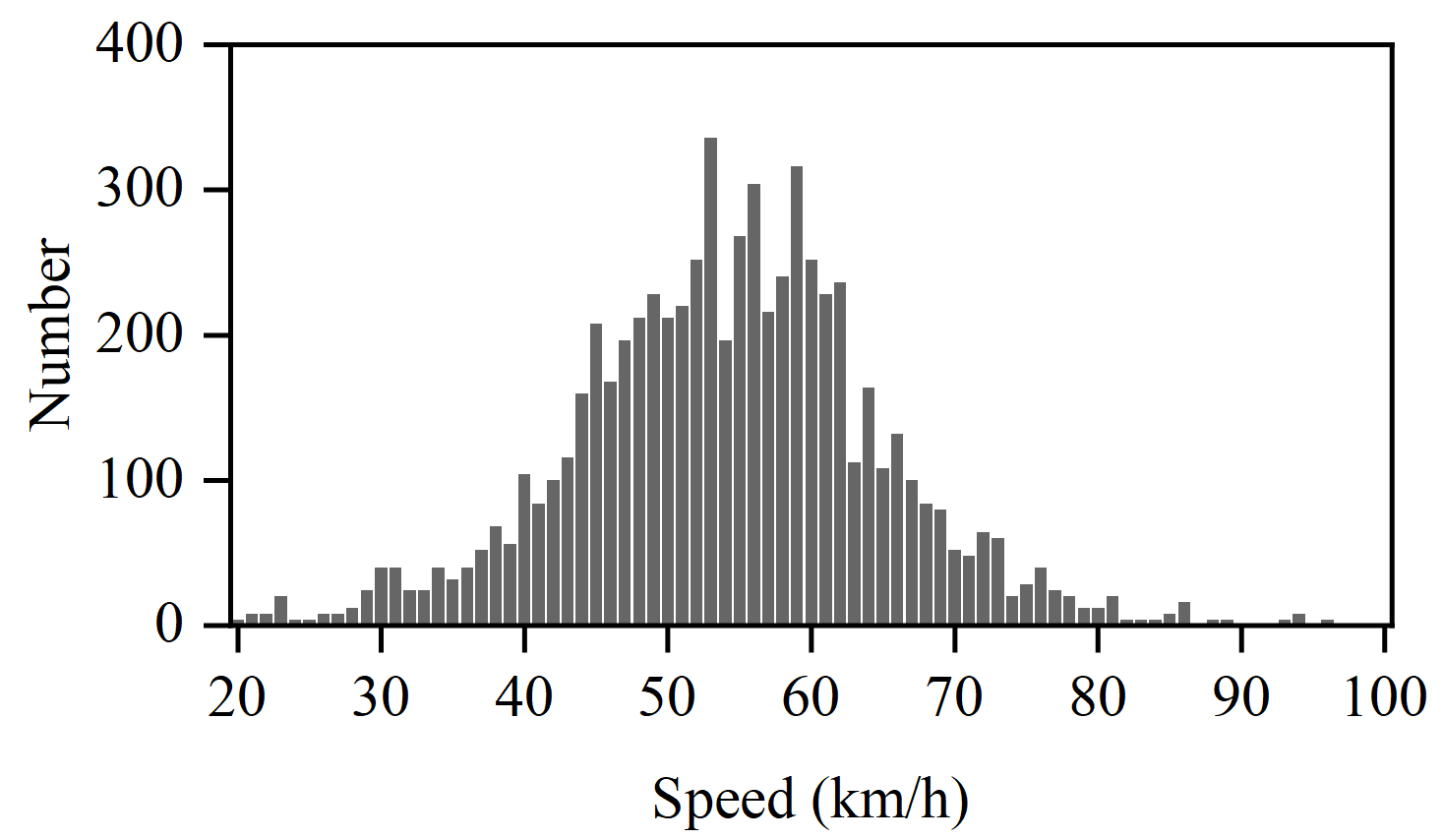}     
        \caption{Statistics on the Number of Vehicles in SZUR-Acoustic Dataset.}
	\label{Number}
\end{figure}

Figure \ref{Number} presents the statistical distribution of vehicle counts at distinct speeds, revealing a pattern that approximates a normal distribution, with the majority of vehicles traveling between 50 $\mathrm{km/h}$ and 60 $\mathrm{km/h}$. Vehicles with speeds below 40 $\mathrm{km/h}$ or above 70 $\mathrm{km/h}$ are comparatively infrequent. Regarding speed classification, the SZUR-Acoustic Dataset maintains consistency with the IDMT-Traffic dataset to enable robust comparative analyses between the two. Specifically, speeds below 45 $\mathrm{km/h}$ are classified as 30 $\mathrm{km/h}$, speeds from 45 $\mathrm{km/h}$ to 60 $\mathrm{km/h}$ are categorized as 50 $\mathrm{km/h}$, and speeds above 60 $\mathrm{km/h}$ are designated as 70 $\mathrm{km/h}$. In total, the study encompasses 4822 audio samples, with respective group sizes of 339 for 30 $\mathrm{km/h}$, 3215 for 50 $\mathrm{km/h}$, and 1268 for 70 $\mathrm{km/h}$. Each audio clip has a standardized duration of 2 seconds.

\subsection{IDMT-Traffic Dataset}
IDMT-Traffic dataset records captured at three distinct locations: Fraunhofer-IDMT (30 $\mathrm{km/h}$), Schleusinger-Allee (50 $\mathrm{km/h}$), and Langewiesener-Strasse (70 $\mathrm{km/h}$) \cite{Dataset}. It includes four types of vehicles: cars, trucks, motorcycles, and buses. This research concentrates exclusively on vehicle speed, categorizing them into three levels: 30 $\mathrm{km/h}$, 50 $\mathrm{km/h}$, and 70 $\mathrm{km/h}$. The study encompasses a total of 17,086 audio samples, with the number of recordings for each speed group being 2,413 for 30 $\mathrm{km/h}$, 9,145 for 50 $\mathrm{km/h}$, and 5,528 for 70 $\mathrm{km/h}$. Each audio clip has a unified length of 2 seconds. 

\subsection{Differences Between IDMT-Traffic and SZUR-Acoustic Datasets}
The IDMT-Traffic dataset and the SZUR-Acoustic dataset exhibit fundamental differences in their data collection methods and the specificity of the recorded data, significantly affecting their utility in various analytical applications. The SZUR-Acoustic dataset is derived from measurements taken on a single roadway, providing a highly controlled environment for capturing speed records. Notably, this dataset does not differentiate between various types of vehicles; instead, it focuses on precise speed measurements that reflect the overall traffic flow on that specific road segment. While this homogeneity in data collection allows for a detailed analysis of speed patterns, it is limited in contextual information regarding vehicle diversity and its effects on traffic dynamics. In contrast, the IDMT-Traffic dataset encompasses a wider array of road environments, thereby allowing for a more comprehensive understanding of vehicle behavior across different contexts. In this dataset, vehicle speeds are estimated based on measurements from various roads, which may lead to reduced precision in the recorded speeds for individual vehicles. Furthermore, the speed classifications drawn from different road measurements may be influenced by varying road conditions, complicating the overall analysis. The reliance on estimations derived from potentially noisy data means that speed approximations are not only contingent on traffic conditions but also affected by diverse road characteristics. Additionally, the IDMT-Traffic dataset features detailed classifications of vehicle types, enriching the analytical framework by enabling comparisons of speed and the impacts of various vehicle categories on traffic dynamics, thereby enhancing the depth of analysis. In summary, while the SZUR-Acoustic dataset offers detailed speed data from a specific location, the IDMT-Traffic dataset provides a broader context through its diverse environmental classifications and comprehensive vehicle type details.
\section{Experimental Details}

\subsection{Evaluation Criteria}
The present study devises a hierarchical, multi-faceted evaluation framework that scrutinises the model from four complementary perspectives: macroscopic performance, statistical agreement, error quantification, and uncertainty characterisation. 

First, conventional classification metrics—including overall \emph{Accuracy}, \emph{Precision}, \emph{Recall} and the $F_{1}$ score are reported to portray discriminative power and misclassification patterns under the prevailing class distribution. 

Second, Cohen’s $\kappa$ coefficient is incorporated to compensate for the influence of random agreement, while the \emph{Mean-Squared Error} (MSE) and \emph{Mean-Absolute Error} (MAE) furnish second- and first-order descriptors of the residuals between predictions and ground truth, thereby quantifying the magnitude and dispersion of predictive bias. 

Third, to assess predictive reliability, the framework integrates modern uncertainty-quantification paradigms: Shannon entropy $H=-\sum_{k}\hat{p}_{k}\log\hat{p}_{k}$ gauges distributional dispersion; the decision-boundary distance $d=\hat{p}_{(1)}-\hat{p}_{(2)}$ identifies samples lying close to the decision surface; and coverage–risk curves are traced across confidence thresholds $\tau\in[0.50,0.95]$ to delineate the model’s risk–benefit profile. 

Fourth, at the sample level, the system logs confidence, entropy, and soft-margin values for every test instance, with particular attention to low-confidence predictions ($\hat{p}_{\max}<0.70$) and high-entropy cases (top decile), thereby exposing the error characteristics of challenging inputs. Simultaneously, the framework continuously monitors the accuracy and loss gaps between the training and validation sets—differences exceeding $0.10$ denote severe over-fitting, whereas values in the $0.05$–$0.10$ range indicate moderate over-fitting—providing an objective basis for subsequent regularisation refinement. 

Finally, class-pair confusion heat-maps and error-distribution plots are generated, pinpointing the most confusable category combinations and furnishing precise guidance for model enhancement. Collectively, this evaluation protocol not only embraces traditional performance indices but also foregrounds predictive confidence, statistical concordance, and over-fitting diagnostics, thereby satisfying the stringent reliability and interpretability demands of high-risk deployment scenarios.

\subsection{Parameter Settings}
This study employs a deliberately tuned deep-learning configuration that balances stability, regularisation, and exhaustive evaluation. During preprocessing a numerical stabiliser $\varepsilon = 1 \times 10^{-8}$ is introduced to safeguard $z$-score standardisation and min–max scaling. The network itself is a dual-branch CNN in which the MFCC and wavelet streams share an identical $48 \rightarrow 96 \rightarrow 192$ filter escalator, each convolution using $3 \times 3$ kernels followed by $2 \times 2$ max-pooling; dropout increases from $0.15$ to $0.25$ across successive blocks, acknowledging the heightened regularisation needs of deeper layers. The shared fully connected head adopts a pyramidal $512 \rightarrow 256 \rightarrow 128$ topology with dropout confined to $0.20$–$0.30$, while an $L_{2}$ penalty of $0.005$ strikes an explicit compromise between over-fitting control and expressive capacity. Training proceeds for $250$ epochs with a batch size of $48$ under Adam ($\beta_{1} = 0.9$, $\beta_{2} = 0.999$) and a conservative initial learning rate of $5 \times 10^{-4}$; a ReduceLROnPlateau schedule (factor $=0.8$, patience $=8$) adaptively reduces the step size when validation loss plateaus. Data are partitioned via an $80{:}20$ stratified split, with $20\%$ of the training portion reserved for validation to preserve class priors throughout.
\section{Results}

\subsection{Results of Speed Classification}

Table \ref{Rsults} reports the classifier’s performance stratified by vehicular speed. At 30 km/h, the model attains only moderate discrimination (precision = 73.44\%, recall = 71.24\%, F1 = 72.31\%; n = 66), a result likely driven by both the low sample count and the reduced signal‐to‐noise ratio characteristic of slow‐moving recordings. In the 50 km/h cohort—the largest group (n = 645)—performance peaks (precision = 87.36\%, recall = 95.35\%, F1 = 91.18\%), indicating that the learned features most effectively capture pacemaker signatures under moderate motion. Remarkably, even at 70 km/h, where motion artifacts are maximal, the classifier remains robust (precision = 92.89\%, recall = 92.05\%, F1 = 91.15\%; n = 254), underscoring its resilience to signal distortion. Overall, the pronounced performance deficit at low speeds highlights the need for targeted preprocessing and data‐augmentation strategies—such as adaptive noise filtering and oversampling of slow‐speed ECG segments—to bolster detection accuracy across all driving conditions.

\begin{table}[!ht]
    \centering
    \caption{Accuracy of Pacemaker Classifiers by Manufacturer of Test Dataset.}
    \label{Rsults}
    \resizebox{1\linewidth}{!}{ 
    \begin{tabular}{cccccccc}
       \hline 
       Speed Range  &Precision (\%) &Recall (\%) &	F1-Score (\%) &	Support  \\
       \hline
       Low Speed (30km/h)    &73.44  &71.24  &	72.31  & 66           \\ 
       Middle Speed (50km/h) &87.36  &95.35  &	91.18  & 645  \\
       High Speed (70km/h)   &92.89  &72.05  &	81.15  & 254 \\
       Total                 &87.56  &   -   &  -     &965\\
       \hline
    \end{tabular}
    } 
\end{table}

The fuzzy‐boundary analysis reveals that the classifier’s mean predictive confidence is \(\bar c = 0.9807\) (\(\sigma_c = 0.0702\); range \([0.5270,1.0000]\)). Systematically raising the confidence threshold from 0.50 to 0.95 yields a marginal accuracy gain (from 0.8756 to 0.8920) concomitant with a reduction in coverage (from 100\% to 92.12\%), thereby highlighting the intrinsic precision–coverage trade‐off of a high‐confidence‐only decision rule. A distinct subset of 26 observations (2.69\% of the dataset) exhibits confidence below 0.70 and attains only 0.6923 accuracy, underscoring the necessity of human‐in‐the‐loop adjudication for low‐confidence cases. Error‐type profiling indicates that the predominant misclassification occurs for true class 2 labeled as class 1 (70 instances), followed by true class 0 as class 1 (19 instances), true class 1 as class 0 (16 instances), true class 1 as class 2 (14 instances), and true class 2 as class 0 (1 instance).These misclassification errors predominantly manifest as confusions between adjacent speed intervals.  

Entropy‐based uncertainty quantification yields a mean entropy of \(\bar H = 0.0503\) (\(\sigma_H = 0.1432\); range \([0.0005,0.7710]\)), and the highest-decile subset (\(n=97\); \(H \ge 0.1051\)) achieves only 0.6907 accuracy, further justifying manual review of high-entropy predictions. Complementarily, decision‐boundary distance analysis reports a mean distance of \(\bar d = 0.9620\) (\(\sigma_d = 0.1388\); range \([0.0576,0.9999]\)), with the lowest-decile cohort (\(n=97\); \(d \le 0.9596\)) attaining 0.7010 accuracy. The joint application of entropy and distance criteria thus furnishes a principled mechanism for isolating high-risk samples.

In the comprehensive quality assessment, the model attains Cohen’s kappa coefficient \(\kappa = 0.7262\), indicative of substantial inter-rater agreement beyond chance. When treating class labels as continuous targets, the observed mean squared error (\(\mathrm{MSE} = 0.1275\)) and mean absolute error (\(\mathrm{MAE} = 0.1254\)) offer complementary perspectives on residual dispersion. From a classification standpoint, macro-averaged precision, recall, and \(F_1\)-score are 0.8456, 0.7954, and 0.8155, respectively, while the weighted-average \(F_1\)-score reaches 0.8725, collectively attesting to the model’s robust discriminative performance under significant class imbalance.

\subsection{Ablation Experiments}
Experimental results, showing in Figure \ref{Ablation} demonstrate that the Balanced Multi-Modal Convolutional Neural Network (BMCNN) achieves the highest overall accuracy, 87.56\%, among the five evaluated algorithms. The single-modal Mel-Frequency Cepstral Coefficients (MFCC) approach ranks second at 84.87\%, marginally surpassing the Baseline Multi-Modal CNN (Base-MCNN), which attains 84.35\%. Speed-resolved analysis reveals that BMCNN attains accuracies of 73.44\%, 87.36\%, and 92.89\% in the low-, medium-, and high-speed regimes. MFCC records 62.50\%, 85.82\%, and 88.67\% under the same conditions, whereas Base-MCNN yields 63.38\%, 85.92\%, and 86.34\%.
These comparisons indicate that Wavelet features provide the best performance in the low-speed scenario (77.78\%), with all other methods remaining below 74\% in this regime. In the medium-speed regime, BMCNN delivers the highest accuracy (87.36\%), followed closely by Base-MCNN and MFCC, with a gap of less than two percentage points. Under high-speed conditions, BMCNN again leads (92.89\%), while MFCC (88.67\%) occupy the second and third positions, respectively. Overall, multimodal approaches offer clear advantages in the medium- and high-speed regimes, whereas the Wavelet representation excels at low speeds, highlighting the complementary nature of distinct feature sets across different operational speeds.
\begin{table}[!ht]
    \centering
    \caption{Accuracy of Pacemaker Classifiers by Manufacturer of Test Dataset.}
    \label{Ablation}
    \resizebox{1\linewidth}{!}{ 
    \begin{tabular}{cccccccc}
       \hline 
       Algorithm  & Low Speed&Middle Speed&High Speed&	Precision (\%) \\
       \hline
       MFCC   &62.50 &85.82 &	88.67  & 84.87        \\ 
       Wavelet &77.78  &79.22 &	81.10&79.84 \\
       Base-MCNN   &63.38 &85.92 &	86.34 & 84.35\\
       BMCNN        &73.44 &  87.36  & 92.89   &87.56\\
       \hline
    \end{tabular}
    } 
\end{table}

\subsection{Robustness Experiments}
\subsubsection{Gaussian Noise}
Table \ref{tab:gaussian} presents the performance evaluation of the BMCNN model under varying Gaussian noise intensities ($\sigma$ = 0.01-0.05), revealing significant speed-scenario-dependent characteristics. Low-speed scenarios demonstrate the highest sensitivity to noise interference, with performance dramatically declining from the baseline 73.44\% to 42.42\% at $\sigma$=0.05, while high-speed scenarios, though also affected, exhibit relatively modest degradation from 92.89\% to 55.91\%. Most remarkably, middle-speed scenarios show anomalous behavior where noise addition actually enhances performance, improving from 87.36\% to a peak of 96.59\% at $\sigma$ = 0.04. This phenomenon suggests potential overfitting in the original model for middle-speed scenarios, where moderate noise serves as a regularization mechanism, enhancing model generalization capability. In terms of overall precision, BMCNN maintains performance above 85\% under mild noise perturbations ($\sigma$ $\le$ 0.03), demonstrating reasonable noise robustness, though overall precision gradually decreases to 82.07\% as noise intensity increases. These experimental findings not only illuminate the stability characteristics of BMCNN across different operational conditions but also provide valuable insights into the behavioral mechanisms of deep learning models in noisy environments, contributing to our understanding of noise-resilient neural network design and optimization strategies.

\begin{table}[!ht]
    \centering
    \caption{Gaussian noise experiments of BMCNN}
    \label{tab:gaussian}
    \begin{tabular}{ccccc}
        \toprule
        $\sigma$ & Low speed & Middle speed & High speed & Precision (\%)\\
        \midrule
        0.01 & 60.61 & 94.42 & 68.11 & 85.18 \\
        0.02 & 57.58 & 95.66 & 67.72 & 85.70 \\
        0.03 & 54.55 & 96.43 & 66.14 & 85.60 \\
        0.04 & 50.00 & 96.59 & 62.99 & 84.56 \\
        0.05 & 42.42 & 96.43 & 55.91 & 82.07 \\
        BMCNN & 73.44 & 87.36 & 92.89 & 87.56 \\
        \bottomrule
    \end{tabular}
\end{table}

\subsubsection{Temporal Shift Noise}

Table \ref{tab:temporal} presents a comprehensive performance comparison between the BMCNN model and six temporal shift noise variants, where three distinct boundary handling strategies are employed: edge strategy extends boundary values to fill shifted positions, wrap strategy uses circular padding by wrapping values from the opposite boundary, and constant strategy fills shifted positions with predetermined constant values. These strategies represent different approaches to managing temporal discontinuities introduced by artificial time shifts. The comparison clearly highlights the significant advantages of the BMCNN architecture and the detrimental effects of temporal shift noise on model performance. In terms of overall precision, BMCNN achieves the highest performance at 87.56\%, leading the best-performing temporal shift variant 2 wrap (80.93\%) by 6.63 percentage points, a margin that substantially demonstrates the superiority of the original BMCNN architecture. Through detailed analysis across speed scenarios, BMCNN exhibits comprehensive dominance: in low-speed scenarios, BMCNN's 73.44\% performance substantially surpasses all temporal shift variants, with even the best-performing 2 edge (56.06\%) lagging by 17.38 percentage points; high-speed scenarios represent BMCNN's absolute advantage domain, where its 92.89\% performance establishes overwhelming superiority, with all temporal shift variants confined to the 60-82\% range, creating maximum gaps exceeding 30 percentage points, indicating that temporal shift noise severely disrupts temporal feature learning in high-speed scenarios. The only noteworthy exception occurs in middle-speed scenarios, where although BMCNN maintains leadership at 87.36\%, 1 edge (95.97\%) and 2 edge (95.66\%) demonstrate anomalous performance improvements, suggesting that specific temporal shift patterns may inadvertently enhance certain feature representations in middle-speed scenarios, resembling regularization effects. Through horizontal comparison of different temporal shift strategies, wrap strategies demonstrate relatively optimal overall performance (1 wrap: 84.73\%, 2 wrap: 80.93\%), followed by edge strategies (1 edge: 83.73\%, 2 edge: 81.66\%), while constant strategies exhibit the weakest performance (1 constant: 84.77\%, 2 constant: 82.80\%), reflecting the varying degrees of impact that different boundary handling approaches have on maintaining temporal continuity. Overall, these experimental results not only validate the superiority of the original BMCNN architecture in temporal modeling but also reveal the critical importance of temporal alignment in deep learning models, demonstrating that any artificial perturbation to temporal structure significantly impairs model learning effectiveness and generalization capability.

\begin{table}[!ht]
    \centering
    \caption{Temporal shift noise experiments of BMCNN}
    \label{tab:temporal}
    \begin{tabular}{ccccc}
        \toprule
       Temporal Shift & Low speed & Middle speed & High speed & Precision (\%)\\
        \midrule
        1 constant &50.00  &91.32&  77.17& 84.77\\
        2 constant &40.91  &87.60&  81.50& 82.80\\
        1 edge     &54.55  &95.97&60.24 &83.73\\
        2 edge     &56.06  &95.66&52.76 &81.66 \\
        1 wrap     &53.03  &95.81&61.02 &83.73\\
        2 wrap     &53.03  &94.57&53.54 &80.93\\
        BMCNN & 73.44 & 87.36 & 92.89 & 87.56 \\
        \bottomrule
    \end{tabular}
\end{table}

\subsection{Generalization Experiments}
Based on the results of BMCNN generalization experiments presented in Table \ref{generalization}, the SZUR and IDMT algorithms demonstrate distinctly different performance characteristics. The SZUR algorithm achieves optimal performance in low-speed scenarios with a recognition rate of 96.67\%, but exhibits significant performance degradation as speed increases, dropping to 79.50\% at middle speed and 78.47\% at high speed, while maintaining an overall precision of 96.69\%. Conversely, the IDMT dataset displays a pronounced high-speed advantage, progressively improving from 73.44\% at low speed to 87.36\% at middle speed and 92.89\% at high speed, with an overall precision of 87.56\%. This performance disparity stems from their underlying technical principles: SZUR employs fuzzy boundary classification methodology, which effectively handles ambiguous boundary conditions in low-speed scenarios by capturing subtle feature variations through fuzzy logic, however, in high-speed dynamic environments, the rapidly changing characteristics may exceed the adaptive capacity of fuzzy boundary processing; whereas IDMT, based on actual measurement data from three road conditions, utilizes non-fuzzy boundary deterministic classification approaches, where multi-road data fusion provides enriched high-speed driving feature information, enabling superior recognition capability and stability in high-speed scenarios, yet lacking the flexibility of fuzzy processing for complex low-speed situations.

\begin{table}[!ht]
    \centering
    \caption{Generalization experiments of BMCNN.}
    \label{generalization}
    \resizebox{1\linewidth}{!}{ 
    \begin{tabular}{cccccccc}
       \hline 
       Algorithm  & Low Speed&Middle Speed&High Speed&	Precision (\%) \\
       \hline
       IDMT Traffic Dataset   &96.62 &95.00&98.47  & 96.28       \\ 
       SZUR Acoustic Dataset   &73.44 &  87.36  & 92.89   &87.56\\

       \hline
    \end{tabular}
    } 
\end{table}

\section{Conclusion}\label{C}
This study introduces a BMCNN algorithm and, for the first time, constructs and publicly releases the SZUR dataset, comprising 4822 vehicular acoustic recordings acquired under controlled urban conditions in Suzhou. All recordings were captured at the same roadside location and span the full vehicular speed range, with a concentration around 50–60 km/h, thus providing a valuable benchmark for acoustic traffic analysis. The proposed model fuses MFCCs with wavelet-transform features, achieving classification accuracies of 87.56\% on SZUR and 96.28\% on the public IDMT-Traffic dataset, which demonstrates its strong generalizability; we also examine inter-dataset differences and their effects on performance. Under Gaussian noise ($\sigma$ $\leq$ 0.05) and various temporal-shift perturbations, overall accuracy declines by only about 5\%, with each speed class remaining above 80\%, confirming the model’s robustness to complex noise. Future work will focus on augmenting low-speed samples and expanding the dataset to further enhance classification accuracy and generalization in low-speed scenarios.
  
\bibliographystyle{elsarticle-num} 
\bibliography{references}

\end{document}